\documentclass[12pt]{article}
\usepackage[dvips]{graphicx}
\usepackage{longtable}

\def\title{THE ORBITAL AND PHYSICAL PARAMETERS OF THE ECLIPSING}
\def\titletwo{BINARY OW GEMINORUM}
\def\author{By C. Ga{\l}an$^1$, M. Miko{\l}ajewski$^1$, T. Tomov$^1$, D. Kolev$^2$, D. Graczyk$^1$,}
\def\authortwo{A. Majcher$^{1,3}$, J. {\L}. Janowski$^1$ and  M. Cika{\l}a$^{1,4}$}
\def\department{$^1$Uniwersytet Miko{\l}aja Kopernika, Centrum Astronomii, Toru\'n, Polska}
\def\departmenttwo{$^2$ National Astronomical Observatory Rozhen, Institute of Astronomy, Smolyan, Bulgaria}
\def\departmentthree{$^3$ Instytut Problem\'ow J\c{a}drowych im. Andrzeja So{\l}tana Otwock -- \'Swierk, Warszawa, Polska}
\def\departmentfour{$^4$ Obserwatorium Astronomiczne im. T. Banachiewicza, W\c{e}gl\'owka, Polska}
\def\abstract{We present our multicolour photometric data of the primary and
secondary eclipses of OW Gem that took place in 1995, 2002, and 2006, as
well as new radial-velocity data collected since 1993 by R.~F.~Griffin and
A.~Duquennoy.  The Wilson--Devinney code was used for the simultaneous
solution of both photometric and spectroscopic data.  A~complete set of
orbital and physical parameters of the components was obtained.  The pair of
values, eccentricity $e= 0.5286$ and argument of periastron $\omega=
140^{\circ}.73$, give better compatibility of the moment of the secondary
minimum with the observations compared to previous estimates.}

\begin{document}

\centerline{\title}
\centerline{\titletwo}

\bigskip

\centerline{\it\author}
\centerline{\it\authortwo}
\centerline{\it\department}
\centerline{\it\departmenttwo}
\centerline{\it\departmentthree}
\centerline{\it\departmentfour}

\bigskip

{\large \abstract}

\bigskip


\noindent{\it Introduction.}\\

OW Gem is an unusual, long-period eclipsing binary, composed of two evolved supergiant stars. Variability
of the star has been noticed by Kaiser$^1$ during photographic searches of nova stars in March 1988.  
Photoelectric observation soon showed a shallow secondary minimum$^2$ at phase 0.23 in the {\textit V} band,
indicating that the orbit has high eccentricity. It turned out that eclipses were visible on Harvard photographic
plates$^3$ already from 1902. The orbital period (1258.59 days - about 3.45 years) was derived from 11 eclipse
events, which took place before 1992$^4$. Radial velocity data of very good quality were obtained
by Griffin $\&$ Duquennoy$^5$ (in tables: G$\&$D), and used by them for the first reliable analysis of the system.
The presence of eclipses distinguish OW Gem from several known similar systems$^6$, hence we know the exact masses
of the components in this case ($5.5 M_{\odot}$ and $3.8 M_{\odot}$). The object has turned out to be unusually
interesting. Both its components have quite large, but considerably different masses and they are now
simultaneously in the short phase of supergiant evolution. Therefore, the evolutionary status of the system
seems to be in contradiction to the current stellar evolutionary models.\\

Many attempts of modelling the system parameters appeared in recent years as a result of several authors
carrying out good quality multicolour photometric measurements of the eclipses. Derekas {\it et al}.$^7$ have
presented a simple model based on observations of the main minimum at the turn of the years 2001 and 2002, in
which they estimate the temperature and the luminosity of the components. Another group (Terrell {\it et al.}$^8$) has
used Wilson -- Devinney code (hereafter: WD) for modelling and operating on Kaiser {\it et al}.$^9$ photoelectric
observational data together with the only available at that time radial velocity measurements$^5$, and they have
obtained a complete set of parameters. In the same year, we presented a simple model$^{10}$ within the confines
of which we have obtained the inclination of the orbit and the most probable temperature $T_2=4950K$ of the
secondary (cool) component for an assumed value of the primary temperature $T_1 = 7100K$. In that model, the
limb darkening was neglected and stellar fluxes were approximated as blackbodies. This work presents new
multicolour photometry covering one (1995) primary eclipse and three (1995, 2002, 2006) secondary eclipses.
We have independently obtained a complete set of parameters using our own and part of Derekas {\it et al}.$^7$
photometric data. The new - partly not published earlier - radial velocity data (Appendix -- Table~\ref{rvGD})
were used, due to kindness of Roger Griffin which has collected them together with A. Duquennoy during the last
fourteen years. Suggested by Dr R.F. Griffin weight were applied. Additionally three our own radial velocity
measurements (Appendix -- Table~\ref{radTor}) were included for analysis.\\


\noindent{\it Observational photometric data}\\

\begin{figure}[!h]
\begin{center}
\includegraphics[angle=0,width=0.85\textwidth]{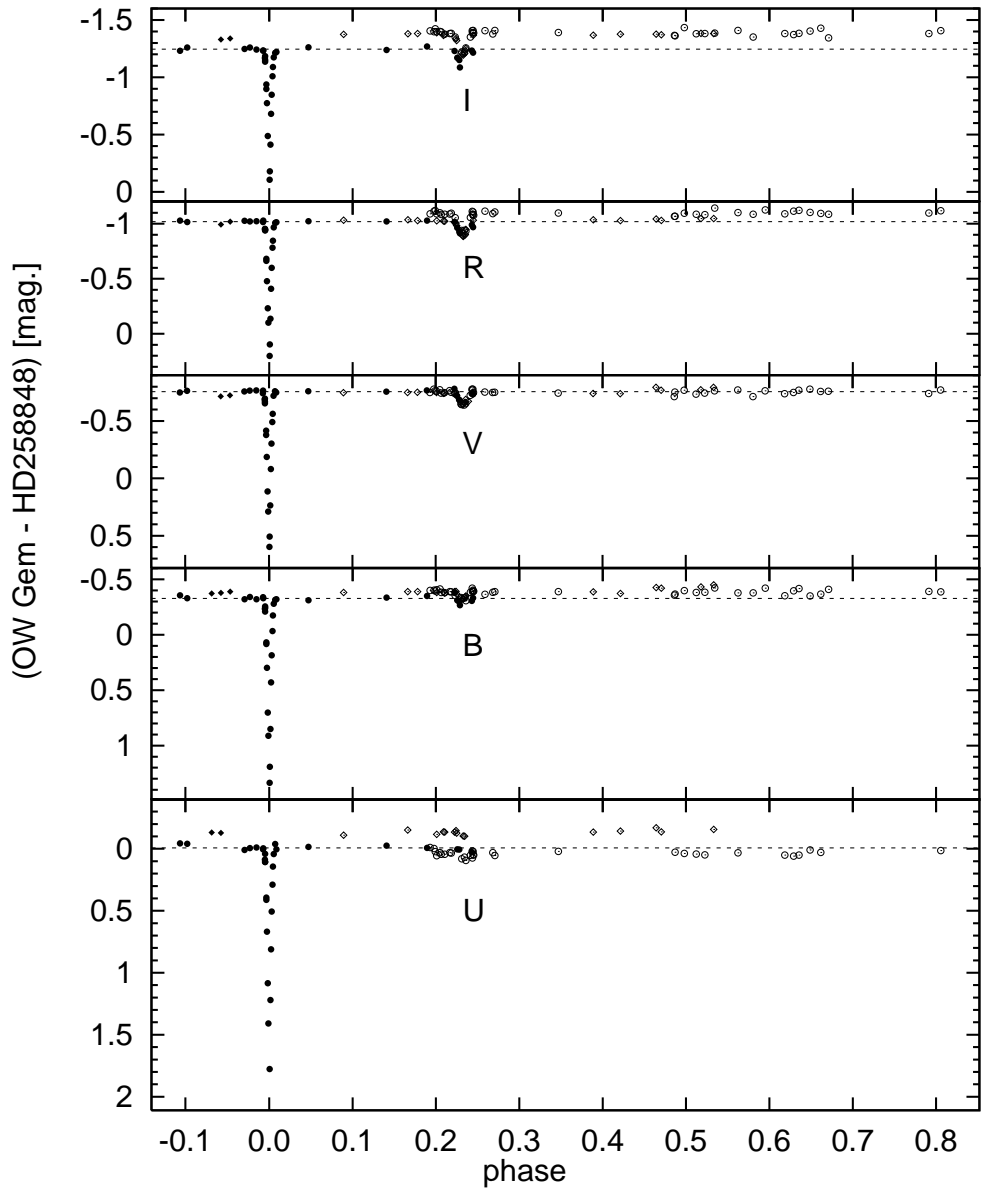}
\caption{\small{The {\it UBVRI} lights curves of OW Gem. Our original differential magnitudes are presented.
The data was phased with a period of 1258.58 days. The horizontal dashed lines mark the average brightness
outside of eclipse in the old photometric system with EMI~9558B photomultiplier. Data collected with
photomultipliers are represented by circles, and these with CCD by diamonds. Filled symbols are used for
EMI~9558B or SBIG:STL~11000 and open for Burle~C31034 or SBIG:STL~1001.}}
\label{lightcurves}
\end{center}
\end{figure}

An international observational campaign was organized by Terrell {\it et al}. during 1995 primary and secondary
eclipses$^{11}$. Responding to this, we obtained multicolour photometry for both events with a 60 cm Cassegrain
reflector at Piwnice Observatory near Toru\'n (Poland). We used a single-channel diaphragm photometer
with non -- cooled EMI~9558B photomultiplier. Our response curves for {\it U}, {\it B} and {\it V} bands were
very close to the standard Johnson's system, whereas our broad {\it R}, {\it I} bands had significantly shorter
mean wavelengths than Johnson {\it RI} and Cousins {\it (RI)$_C$} (Table~\ref{lambsyst}). The accuracy of our
measurements was $\pm 0.03$, $\pm 0.02$, $\pm 0.017$, $\pm 0.02$ and $\pm 0.019$ in {\it UBVRI} respectively.
Unfortunately, in the analysis based on the 1995 campaign$^9$ our photometric data were not included.
Our observations of the 1995 secondary eclipse have poor time covering as a result of bad weather.
To fill the gaps in the light curves, new data were obtained nearby and during the 2002 secondary eclipse.
A single--channel diaphragm photometer, with a cooled Burle~C31034 photomultiplier and a set of five filters
{\it U}, {\it B}, {\it V}, {\it R$_C$}, {\it I$_C$} were used. Their response curves were close to the standard
Johnson--Cousins {\it UBV(RI)$_C$} system (Table~\ref{lambsyst}). Additionally, two intermediate--band interference
filters ($FWHM \approx 100\AA$), $"{\it h}"$ (located at $H_{\beta}$ around $\lambda = 4870 \AA$) and $"{\it c}"$
(located in the continuum around $\lambda~=~4804~\AA$) were used. The accuracy of these measurements was $\pm 0.028$,
$\pm 0.021$, $\pm 0.018$, $\pm 0.017$ and $\pm 0.021$ in {\it UBVR$_C$I$_C$} respectively.
Data around the last 2006 secondary minimum have been obtained with two new CCD detectors (SBIG: STL~11000
and STL~1001) with new filters set. The mean wavelength of these photometric systems are presented and compared
with the two previous photometric systems in Table \ref{lambsyst}. The accuracy was $\pm 0.018$, $\pm 0.007$,
$\pm 0.005$, $\pm 0.007$ and $\pm 0.008$ in {\it UBVR$_C$I$_C$} filters respectively. HD 258848 was chosen as
a comparison star and GSC 1332:0578 as a check star, both suggested by Terrell {\it et al}.$^{11}$. Our original
differential magnitudes ($OW Gem - HD258848$) are presented in the Appendix (Tables \ref{photEMI},\ref{photBurle},
\ref{photSTL11000},\ref{photSTL1001}), and as {\it UBVRI} light curves in Figure~\ref{lightcurves}.

\begin{table}[!h]
\begin{center}
\caption{\small{Mean wavelength of the four photometric systems used by us.}}
\label{lambsyst}
\begin{tabular}{ccccc}
receiver: & EMI 9558B & Burle C31034 & SBIG:STL 11000 & SBIG:STL 1001 \\
band   &  $\bar\lambda [\AA]$ & $\bar\lambda [\AA]$ & $\bar\lambda [\AA]$ & $\bar\lambda [\AA]$ \\
U       & 3708   & 3678 & 3676 & $\sim$3600 \\
B       & 4342   & 4467 & 4392 & $\sim$4400 \\
V       & 5436   & 5426 & 5343 & 5404 \\
R       & 6391   & 6689 & 6319 & 6414 \\
I       & 7420   & 8380 & 8020 & 8305 \\
\end{tabular}
\end{center}
\end{table}


\noindent{\it Period analysis}\\

\begin{figure}[!t]
\begin{center}
\includegraphics[angle=270,width=0.95\textwidth]{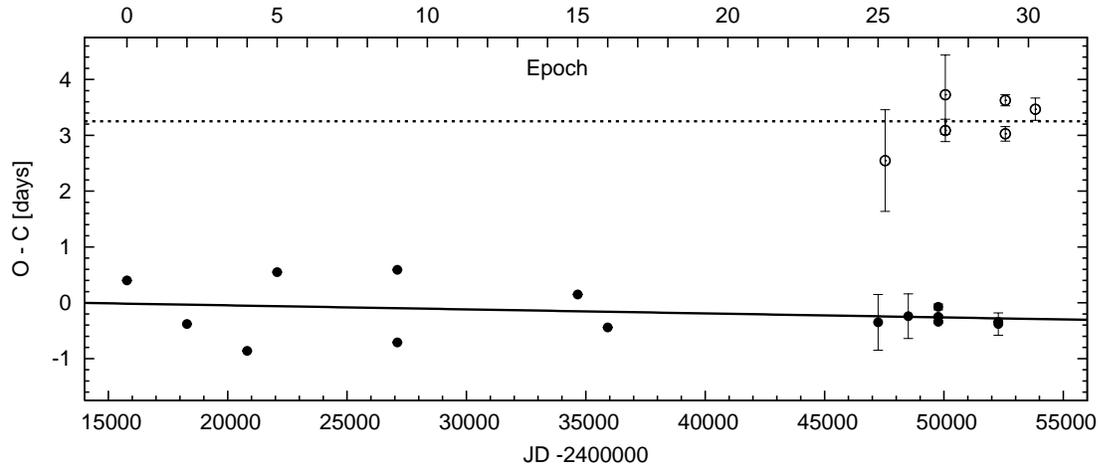}
\caption{\small{The O-C diagram for the moments of the primary eclipse from the ephemeris given by
Equation~\ref{sgefem} (filled circles) and the secondary eclipse from the ephemeris given by Equation~\ref{ngefem}
with phase shift adopted as $\Delta\phi_{II}=0.23$ (open circles). The best fit for the primary eclipses
(solid line) indicates a inconsiderable shorter period. The best fit for the secondary eclipses (horizontal dashed
line) was found assuming the new value P=1258.58.}}
\label{ocow}
\end{center}
\end{figure}

The O-C analysis was carried out for verification of the orbital period and for determination of the time
between primary and secondary minima. The times of minima from our obserwations and the time of the secondary
minimum from Williams$^2$ were obtained using Kwee and Van Woerden method$^{12}$. The times of collected
minima are presented in Table~\ref{tabmin}. The moment of primary minimum with $E=18$ was excluded from
the analysis because of its obviously high error. The values of O-C residuals for the primary events were
calculated from Williams $\&$ Kaiser's$^4$ ephemeris:

\begin{center}
\begin{equation}
JD_{min I}= 2415779.0 (\pm 0.4) + 1258.59 (\pm 0.03) \times E, 	\label{sgefem}
\end{equation}
\end{center}

\noindent and are shown in Figure \ref{ocow}. The best fit to these data gives the new ephemeris: 

\begin{center}
\begin{equation}
JD_{min I}= 2415778.98 (\pm 0.22) + 1258.580 (\pm 0.011) \times E \label{ngefem}
\end{equation}
\end{center}

\newpage

\begin{table}[!h]
\begin{center}
\caption{\small{The moments of primary and secondary eclipses of OW Gem used for O-C analysis.
The values of O-C for the primary were derived using equation \ref{sgefem}, and for the secondary
with the adopted initial value of the phase shift $\Delta\phi_{II}=0.23$.}}
\label{tabmin}
\begin{tabular}{llllrclc}
&&&&&&&\\
E       & JD-2400000   &   error       &~~~~~~~~~& O-C     &~~~&  Author       & Ref \\
\multicolumn{8}{c}{\bf{Primary}}\\
$0 $      & $15779.4$      &  --           &&~~$0.400$     &   &  Kaiser                & ~4  \\
$2 $      & $18295.8$      &  --           &&$-0.380$      &   &  "                     & ~4  \\
$4 $      & $20812.5$      &  --           &&$-0.860$      &   &  "                     & ~4  \\
$5 $      & $22072.5$      &  --           &&~~$0.550$     &   &  "                     & ~4  \\
$9 $      & $27105.6$      &  --           &&$-0.710$      &   &  "                     & ~4  \\
$9 $      & $27106.9$      &  --           &&~~$0.590$     &   &  "                     & ~4  \\
$15$      & $34658.0$      &  --           &&~~$0.150$     &   &  Fuhrmann              & ~4  \\
$16$      & $35916.0$      &  --           &&$-0.440$      &   &  "                     & ~4  \\
$18$      & $38435.0^\star$& --            &&$-1.380^\star$&   &  "                     & ~4  \\
$25$      & $47243.4  $    & $\pm0.5  $    &&$-0.350$      &   &  Kaiser {\it et al}.   & ~4  \\
$26$      & $48502.1  $    & $\pm0.4  $    &&$-0.240$      &   &  Williams $\&$ Kaiser  & ~4  \\
$27$      & $49760.857$    & $\pm0.052$    &&$-0.073$      &   &  Hager                 & 13  \\
$27$      & $49760.68 $    & $\pm0.03 $    &&$-0.250$      &   &  this work             & --  \\
$27$      & $49760.59 $    & $\pm0.02 $    &&$-0.340$      &   &  Kaiser {\it et al}.   & ~9  \\
$29$      & $52277.77 $    & $\pm0.01 $    &&$-0.340$      &   &  Kaiser {\it et al}.   & ~9  \\
$29$	& 52$277.73$     & $\pm0.2$        &&$-0.380$      &   &  Derekas {\it et al}.	& ~7  \\
\multicolumn{8}{c}{\bf{Secondary}}\\
25      & $47535.50$     & $\pm0.91$       &&~~$2.547$     &   &  Williams              & ~2  \\
27      & $50053.84$     & $\pm0.71$       &&~~$3.727$     &   &  this work             & --  \\
27      & $50053.2 $     & $\pm0.2 $       &&~~$3.087$     &   &  Kaiser {\it et al}.   & ~9  \\
29      & $52570.9 $     & $\pm0.1 $       &&~~$3.627$     &   &  Kaiser {\it et al}.   & ~9  \\
29      & $52570.30$     & $\pm0.13$       &&~~$3.027$     &   &  this work             & --  \\
30      & $53829.32$     & $\pm0.20$       &&~~$3.467$     &   &  this work             & --  \\
\multicolumn{8}{l}{$\star$ - {\it excluded from our analysis}}
\end{tabular}
\end{center}
\end{table}

\noindent The values O-C for the secondary minima were calculated assuming a new period P=1258.58 days
(Equation~\ref{ngefem}) and taking into acconut an initial value of the phase shift $\Delta\phi_{II}=0.23$
according to Williams$^2$ which corresponds to 289.473 days (Figure~\ref{ocow}). The residuals obtained
for the six measured secondary minima give a mean value ${\overline{O-C}}$=3.25 $\pm$0.19 days and hence
the new phase shift is $\Delta\phi_{II} = 0.23258 \pm0.00016$.\\

\newpage

\noindent{\it Preparation of photometric data}\\

We have used five sets of photometric data for modelling. Four sets were obtained in Piwnice Observatory, two with
photomultiplier detectors and two with CCDs. The fifth set of CCD data were obtained by Derekas {\it et al}.$^7$.
They reported a possible existence of asymmetry in the light curve of the primary eclipse. This effect is not
visible in any other data. It seems to be an artefact. Figure~1 in Derekas {\it et al}.$^7$ paper shows that all
of the {\it VR$_C$I$_C$} Szeged Observatory data points lie somewhat above the fit to Piszk\'estet\"o Observatory
data points. Because of this we used only Piszk\'estet\"o data to improve the primary eclipse time covering.
As a consequence we have collected data from five different photometric systems, which are somewhat shifted
in magnitude. Additionally, the depth of minima in particular bands depends on their mean wavelengths.
The correction for the depth of secondary eclipse is very small in respect to primary eclipse.
To transform the data to the homogenous systems without correction for depth of primary eclipse, we
shifted the {\it BVR$_C$} Piszk\'estet\"o CCD data (the star GSC 1332-0578 was the comparison star for this data),
and UBVR Toru\'n data obtained with Burle~C31034 photomultiplier and STL CCDs to the data obtained in 1995 with
EMI~9558B photomultiplier (Table~\ref{lambsyst}).
The {\it I} passband established by EMI~9558B photomultiplier differs considerably from both Johnsons {\it I}
and Cousins {\it I$_C$}, so the {\it I} passband data obtained in the region of phase of primary eclipse were excluded
from analysis. The {\it I} passband established by STL 1001 CCD detector is quite close to the Cousins {\it I$_C$} passband,
so this system was adopted as the reference system for the infrared domain.
The magnitudes from transformed systems were shifted on reference systems by values presented in Table \ref{shftv}.

\begin{table}[!h]
\begin{center}
\caption{\small{Values of shifts from transformed systems on reference systems.}}
\label{shftv}
\begin{tabular}{lccccccc}
                &~~~~&\multicolumn{4}{c}{Ref.syst.}              &~~~~& Ref.syst.\\
                &    &\multicolumn{4}{c}{EMI9558B}               &    & STL1001  \\
                &    & $ U    $ & $ B    $ & $ V    $ & $ R    $ &    & $ I    $ \\
Trans.syst.     &    &          &          &          &          &    &          \\
EMI9558B        &    &  --      &  --      &  --      &  --      &    & $-0.134$ \\
Burle C31034    &    & $-0.036$ & $+0.052$ & $-0.008$ & $+0.076$ &    & $+0.015$ \\
STL 11000       &    & $+0.128$ & $+0.044$ & $-0.041$ & $-0.014$ &    & $-0.041$ \\
STL 1001        &    & $+0.137$ & $+0.062$ & $-0.005$ & $+0.013$ &    &  --      \\
Piszk\'estet\"o &    &  --      & $+0.979$ & $+0.891$ & $+0.860$ &    &  --      \\
\end{tabular}
\end{center}
\end{table}

\noindent Additionally, the depths of the secondary minima in transformed systems were corrected
on reference systems according to expresion:

\begin{center}
\begin{equation}
m_{ref.} = \overline m_{ref.} + (m_{trans.} - \overline m_{trans.}) \times \alpha_{ref./trans.}
\end{equation}
\end{center}

\noindent where individual values of measured brightnesses are denoted by $m$, mean brightnesses outside of
eclipses by $\overline m$ (see~Table~\ref{meanbr}), and the $\alpha $ parameter is ratio of the depth minimum
in reference systems to the depth minimum in transformed systems (see~Table~\ref{alphv}).

\begin{table}[!h]
\begin{center}
\caption{\small{Mean differential magnitude (OW Gem - HD 258848) outside the eclipses in particular photometric
systems.}}
\label{meanbr}
\begin{tabular}{lrrrrr}
             & $ U    $ & $ B    $ & $ V    $ & $ R    $ & $ I    $ \\
EMI9558B     & $-0.002$ & $-0.336$ & $-0.762$ & $-1.020$ & $-1.241$ \\
Burle C31034 & $ 0.034$ & $-0.388$ & $-0.754$ & $-1.096$ & $-1.390$ \\
STL 11000    & $-0.130$ & $-0.380$ & $-0.721$ & $-1.006$ & $-1.334$ \\
STL 1001     & $-0.139$ & $-0.398$ & $-0.757$ & $-1.033$ & $-1.375$ \\
\end{tabular}
\end{center}
\end{table}

\noindent Values of the $\alpha$ parameter were
obtained by fitting a 2nd order polynomial $Depth(\overline \lambda)$ to the observational depths of the
secondary eclipse data obtained with the Burle~C31034 photomultiplier. All the brightnesses were normalized
to unity corresponding to the mean magnitudes outside the eclipse (Table~\ref{meanbr}) as it is needed by
the differential correction (DC) procedure of the Wilson -- Devinney code.\\

\begin{table}[!h]
\begin{center}
\caption{\small{Ratio ($\alpha_{ref./trans.}$) of the depth secondary minima in reference systems to the
depth secondary minima in transformed systems.}}
\label{alphv}
\begin{tabular}{lcccclc}
$\alpha_{ref./trans.}$              & $U    $ & $B    $ & $V    $ & $R    $& $\alpha_{ref./trans.}$         & $I    $ \\
$\alpha_{EMI9558B/BurleC31034}$     & $1.019$ & $0.951$ & $1.002$ & $0.954$& $\alpha_{STL1001/EMI9558B}$    & $1.087$ \\
$\alpha_{EMI9558B/STL1001}$         & $1.020$ & $0.980$ & $1.009$ & $0.996$& $\alpha_{STL1001/BurleC31034}$ & $0.994$ \\
\end{tabular}
\end{center}
\end{table}


\noindent{\it Modelling the system parameters - methods of solution}\\

We have calculated a simultaneous solution to the photometric data described above and the velocity curves. We
have used both the DC and light and velocity curve (LC) programs of the 2003 version of the Wilson~--~Devinney
code$^{14,15}$, where the radiative functions used are based on Kurucz's stellar atmosphere models.
This allows us to model giants or supergiants in addition to main sequence stars. So, the use of this version
of the WD program is more suitable in the OW~Gem case. On the other hand in the 2003 version the previous
effective wavelength characterization of the bandpasses was replaced by integration over the actual bandpasses
of the standard photometric systems. Twenty five standard bands are available in the code. Five of our bands
\textit{UBVRI} used in the analyses adjust well for the \textit{UBVR$_C$I$_C$} standard bands defined in the
program. Unfortunately, our {\textit{h,c}} bands do not have equivalents among the standard ones, so we
decided to omit them in our analyses.

\noindent The Levenberg -- Marquardt's algorithm used in the 2003 version of the WD program together with a
properly selected value of the $\lambda$ parameter (see e.g. Kallrath $\&$ Milone$^{16}$; Kallrath {\it et al}.$^{17}$)
usually allows finding a solution in the parameter space even when a large group of free parameters for
simultaneous iterations is used. However, the OW Gem orbit is characterized by strong eccentricity which
amplified correlations between the parameters. We have used the method of multiple subsets (MMS) recommended
by Wilson~$\&$~Bierman$^{18}$ for dealing with this problem. The method relies on disposing the strongest
correlation through separation of the most correlated parameters into different groups. In our case the groups
have included two (seldom three) weakly correlated parameters. The use of the method does not significantly
extend the time of calculation, but a problem appears in the form of unrealistically small values of the
errors. We have received more plausible values of errors by executing additional iterations with all the free
parameters simultaneously. The solution was calculated with the aid of a program we wrote for "semiautomatic"
iteration, which made it possible to keep control over the result of each iteration by visual inspection of the
evolution of parameters, errors and residuals on the computer screen as a function of the iteration number.
The criterion for a proper solution was set as obtaining the minimum sum of squared residuals over the domain
of parameter space as well as (in some cases) visual inspections of the shape of the $\chi^2$ surface.\\

\noindent{\it Fixed parameters}\\

OW Gem photometric behaviour does not indicate the possibility of the occurrence of spots on the component's
surface, so the system was treated as not spotted. Synchronous rotation has been assumed. It is impossible to
find out the temperatures of both the components in OW Gem case using only the WD code, so the temperature
$T_1$ of the hot component was fixed at $7100 K$ according to the $F2 Ib-II$ classification$^5$. The values of
both stars' temperatures are below $7200~K$, which is a theoretical upper threshold for convective envelopes.
Such cases are characterized by the theoretical values of a bolometric albedo $A=0.5$ and the exponent in
the bolometric gravity brightening $g=0.32$ (Lucy$^{19}$). The third light contribution has been neglected.
The derivatives of the orbital period and $\omega$ were assumed to be equal to zero. A nonlinear, logarithmic
limb darkening law was used. The theoretical coefficients {\it x,y} have been calculated according to
the Van Hamme$^{20}$ tables for: $T_1=7100K$, $T_2=4950K$, log $g_1=2.2$, log $g_2=2.0$.\\

\noindent{\it The free parameters' initial values}\\

The initial values of $a \sin{i} = 1052 R_{\odot}$, $V_{\gamma} = -5.25 km s^{-1}$ and $q=M_2 M_1^{-1}=0.676$
have been adopted from Griffin $\&$ Duquennoy$^5$. The values of the surface linear potentials of the stars
were initially estimated as $\Omega_1=35$, $\Omega_2=24$ from the value of the radius $R_1=30R_{\odot}$,
$R_2=35R_{\odot}$, and mass ratio. A temperature $T_2=4950 K$ and an inclination $i=89^{\circ}$ have been
taken from our simple model$^{10}$. The large orbital eccentricity enables to estimation of the parameters $e$
and $\omega$ when we have information about the time separation of the eclipses and about the duration of the
phenomena. From there we have found the initial values of $e=0.5183$ and $\omega=144.04^{\circ}$ using formulae
4.4.60 and 4.4.61 from Kallrath $\&$ Milone$^{16}$. The parameter $\phi_0$ was treated differently. This parameter
is connected with manner of orbit solution and it is only formal parameter of WD code, where for circular orbits
phase of periastron is equal to 0.0 for $\omega=90^{\circ}$ by definition. If we want to get actual value of
periastron phase with WD code for eccentric orbit we have to take into account the value of phase correction $\phi_0$.
This parameter can be briefly defined as difference of actual periastron phase for eccentric orbit and periastron
phase that would be for circular orbit with adopted the same periastron longitude $\omega$.
Three parameters: eccentricity $e$, periastron longitude $\omega$ and $\phi_0$ parameter are each other dependent.
For that reason parameter $\phi_0$ can not be treated directly as the other free parameters, but it have to be
searched through the wide area of this parameter's space in order to find the global minimum and not to land in
a local one.\\


\noindent{\it The orbital geometry solution}\\

\begin{figure}[!b]
\begin{center}
\includegraphics[angle=270,width=0.9\textwidth]{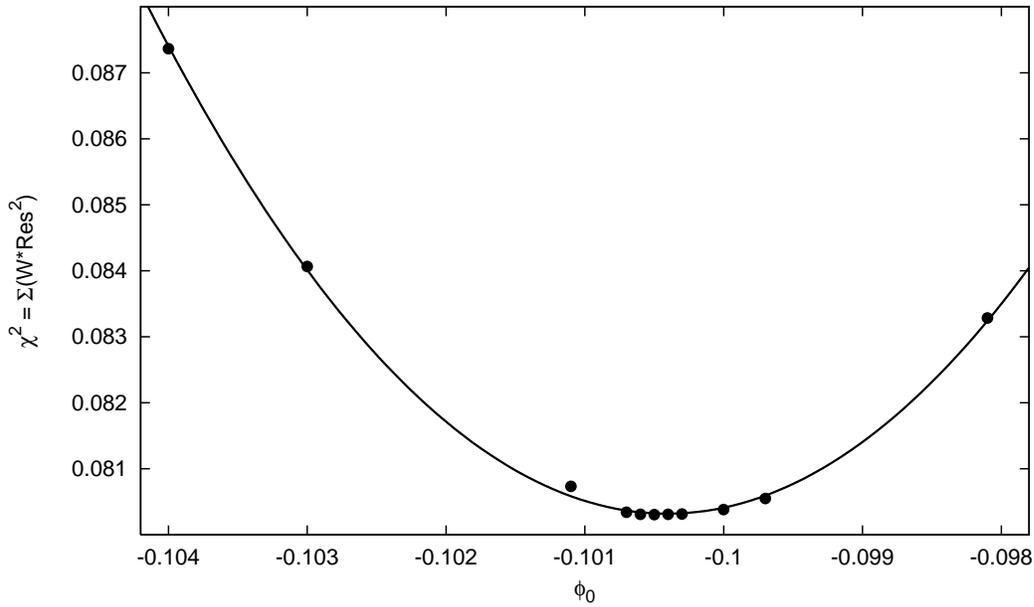}
\caption{\small{Quality of fit obtained for different values of $\phi_0$ parameter.
The polynomial f(x) was fitted and the minimum of the function was found.}}
\label{Sresfi0}
\end{center}
\end{figure}

\begin{figure}[!t]
\begin{center}
\includegraphics[angle=270,width=0.65\textwidth]{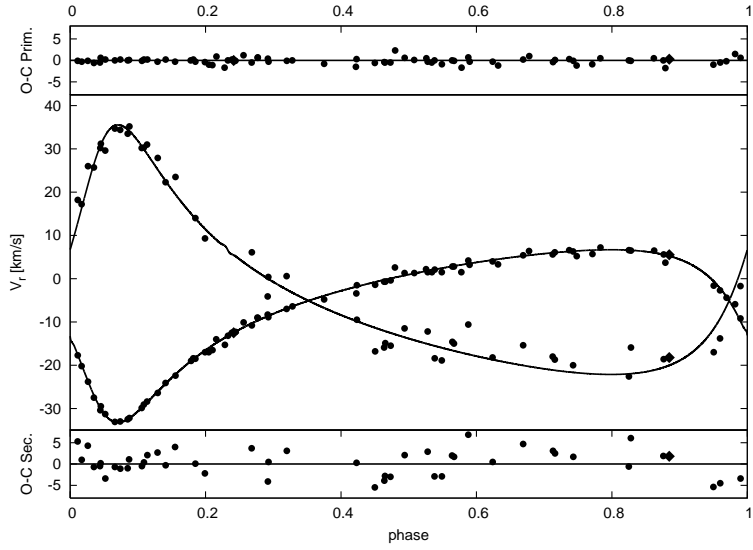}
\caption{\small{Synthetic radial velocity curves (lines) fitted to the Griffin $\&$ Duquennoy
data represented by circles and the our three points represented by diamonds. At the top and the bottom
the residuals (O-C) are placed for the primary nad the secondary components respectively, which
demonstrate the quality of the fit.}}
\label{syntradialcb}
\end{center}
\end{figure}

The solution was carried out in two basic steps. Both stages of calculations were carried out for many values
of $\phi_0$ parameter according to described below procedure. The first stage aimed to determine the geometry
of the orbit through estimation of the parameters: $a \sin{i}$, $e$, $\omega$, $V_{\gamma}$, $q$ and $L_1$. The
bandpass luminosity $L_1$ of the primary component is defined in details in the manual of the 2003 version of
the WD program. The changes in two parameters, the argument of the periastron $\omega$ and the eccentricity $e$,
have a strong influence on the light curves as well as on the velocity curves. The duration of both eclipses and
their phase shift depend very strongly on the variations of these two parameters. This timing puts a strong
limitation on the values $e$ and $\omega$. The WD program enables solving both velocity curves and many light
curves simultaneously$^{21}$.  This advantage of WD code has been used in the first part of the first stage,
where the values of the parameters $e$, $\omega$, $L_1$ were corrected. When a convergence was achieved, then
in the second part of the first stage, the parameters $a \sin{i}$, $V_{\gamma}$, $q$, which depend only on
velocity curves were corrected, then both steps were repeated. The second stage had the purpose of determining
first of all the orbital inclination $i$ together with the other parameters depending only on the light curves:
$T_2$, $\Omega_1$, $\Omega_2$, $L_1$. $\Omega_1$ and $\Omega_2$ are linear functions of the true potentials on
the equipotential surfaces of the stars$^{14}$. A black--body approximation was used. In order to make the
geometric solutions independent of the input values of the radii and the inclination, the first stage of the
solution (searching for $a sin{i}$, $e$, $\omega$, $V_{\gamma}$, $q$) has been repeated again. A further repeating
of both stages did not show changes in the errors limits, so it was considered that for the current value of
$\phi_0$ parameter the final solution for the orbital parameters has been achieved. Each value of $\phi_0$
parameter relate to one value of sum of weighted squared residuals ($\chi^2 = \Sigma(W \cdot Res^2)$), which
constitute the quality of obtained fit. The second order polynomial was fitted for $\chi^2(\phi_0)$ function
and the minimum of the function was found (Figure \ref{Sresfi0}). This way the final set of orbital parameters
has been found and these values are compared with previous solutions in Table \ref{parorb}.
The radial velocity curves computed from our parameters (Figure \ref{syntradialcb}) differ slightly from those
published earlier by Griffin~$\&$~Duquennoy$^5$ and Terrell {\it et al}.$^8$. A good timing of both minima gives
better values for $\omega$ and $e$ in comparison to previous solutions. However, these parameters force small
changes in $a sin{i}$, $V_{\gamma}$ and $q$ in comparison to the free fitting.

\begin{table}[!h]
\begin{center}
\caption{\small{Orbital parameters of OW Gem.}}
\label{parorb}
{\small
\begin{tabular}{llllll}
                 &This work            & Griffin 2007$^{\star \star}$      &Terrell {\it et al}.$^8$ & G$\&$D$^5$       &unit\\
$P$              &$1258.58$            &$1259.30$          &$1258.59$                &$1260.00$         &day\\
$a \sin{i}$      &$1030.0  \pm 10.0$   &$1035.9 \pm 11.8$  &$1044.4 \pm 8.8$         &$1052.0 \pm 18.7$ &$R_{\odot}$\\
$i$              &$89.040 \pm 0.028$   &--                 &$89.09 \pm 0.02$         &$89.0 \pm 0.1$    &degree\\
$V_{\gamma}$     &$-5.10  \pm 0.10$    &$-5.21 \pm 0.06$   &$-5.18 \pm 0.14$         &$-5.25 \pm 0.16$  &$km s^{-1}$\\
$q$              &$0.692 \pm 0.011$    &$0.687 \pm 0.013$  &$0.664 \pm 0.002$        &$0.676 \pm 0.014$ &--\\
$e$              &$0.5286 \pm 0.0006$  &$0.5233 \pm 0.028$ &$0.51718 \pm 0.00002$    &$0.515 \pm 0.011$ &--\\
$\omega$         &$140.73 \pm 0.12$    &$140.3 \pm 0.5$    &$143.08 \pm 0.02$        &$140.2 \pm 1.3$   &degree\\
$\phi_0$         &$-0.1004 \pm 0.0001$ &--                 &$-0.1030 \pm 0.0001$     &--                &--\\
$\Delta\phi_{II}$&$0.23250$            &$0.23655$          &$0.23207$                &$0.24123$         &--\\
$\delta\phi_{II}$$^\star$      & $+0.10$             &$-5.00$            &$+0.64$                  &$-10.89$          &day\\
\multicolumn{6}{l}{$\star$ - {\it The differences between the observed and calculated phase shift of the secondary eclipse}}\\
\multicolumn{6}{l}{$\star \star$ - {\it Private communication}}
\end{tabular}
}
\end{center}
\end{table}

\noindent The final solution must give a formally larger standard deviation of the observational points than a free
fitting to the radial velocities only, without any timing constraints. This is in contradiction to the Terrell
{\it et al}.$^8$ solution, who obtained unrealistically low errors for $e$ and $\omega$, as commented by Griffin$^{22}$.
Moreover, a detailed inspection

\newpage

\noindent of "Figure~1" in Terrell {\it et al}.$^8$ shows, that almost all the points which
lie on the descending branch of the secondary minimum are above their synthetic light curves while, the points
which lie on the ascending branch are located below their model. This disagreement in the timing has been
previously noted by Griffin$^{22}$. Values $e$ and $\omega$ parameters obtained by Terrell {\it et al}.$^8$
differ significantly from the others results and we have found that their solutions must landed in the local minimum
as a consequence of use incorrect $\phi_0$ value. Values $e$ and $\omega$ parameters  which Griffin (2007, private
communication) has obtained with his own, new radial velocity (Appendix -- Table \ref{rvGD}) are close and almost
consistent in borders of errors to our values. However, becuse of lack of photometry in those solution it is not proper
pair of values, and such $e$ and $\omega$ parameters can not give a good timing of the secondary minimum.
The differences between the observed phase shift of the secondary eclipse $\Delta\phi_{II}=0.23258$ and that derived
from the orbital solution (Table~\ref{parorb}) are $+2.4$ hours for our solution, $+15.4$ hours for Terrell
{\it et al}.$^8$ and minus a few days for orbital parameters carried out from radial velocity only. Taking into
account that we have obtained such a good timing from our analysis, we hope that our parameters are close to the
true ones with realistic errors.\\


\noindent{\it System components physical parameters solution}\\

Knowledge about the orbital geometry allows us to proceed with a part of the modelling leading to exact
information about the physical parameters of the components, i.e. temperatures, radii, masses, luminosities.
At this stage of the solution a stellar atmospheres approach has replaced the previous black-body approximation.
The temperature of the hot component have to be adopted. It is not possible to determine temperature of the both
OW Gem components if we do not have at disposal very good quality photometry (accuracy better than $0.^m01$)
of both eclipses reaching deep UV and far infrared domain. We have not, and we can obtain temperatures ratio only.
The Griffin $\&$ Duquennoy$^5$ have classified the hot component spectral class as $F2Ib-II$, that according to
Strai\v{z}ys~$\&$~Kuriliene$^{23}$ spectral class - effective temperature classification gives value of temperature
$T_1=7100K$. This value of hot component temperture was adopted in our calculations, the same as in other papers.
However, we have compared of the OW Gem spectrum with spectra of neighbour class standards, and we have found that
the accuracy of this classification is of the order of one subclass. By interpolations with
Strai\v{z}ys~$\&$~Kuriliene$^{23}$ spectral class - effective temperature classification we have estimated
uncertainity for temperature of hot component $T_1=7100$$^{+150}_{-200}K$ and for corresponding temperature of cold
component via temperature ratio as $T_2=4975$$^{+110}_{-140}K$ (see table \ref{parfot}). The error of our $T_2$ value
given in table \ref{parfot} it is error of fit to observational data and not the error of parameter determination.

\newpage

\begin{figure}[!b]
\begin{center}
\includegraphics[angle=270,width=0.7\textwidth]{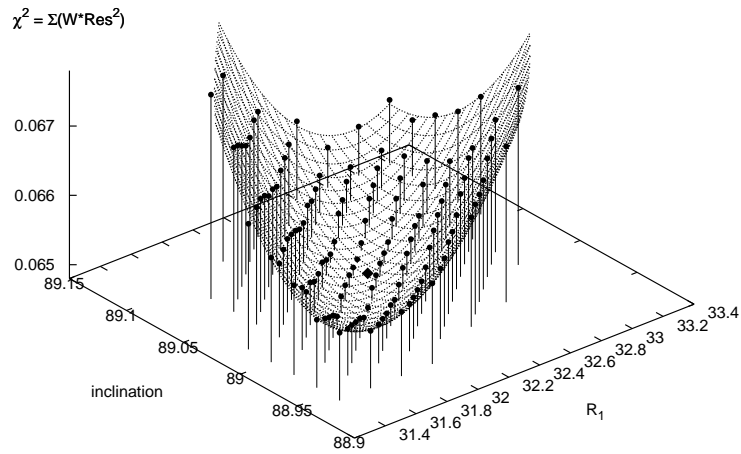}
\caption{\small{Adjusting of the polynomial $f(i,R_1)$ for the points on the $\chi^2$ surface.
The place of the minimum on this surface is denoted by a diamond.
The radii of primary component ($R_1$) is expressed in solar radius units and inclination in degrees.}}
\label{SIEC3D_CHI_Wcb}
\end{center}
\end{figure}

\begin{figure}[!t]
\begin{center}
\includegraphics[angle=270,width=1.0\textwidth]{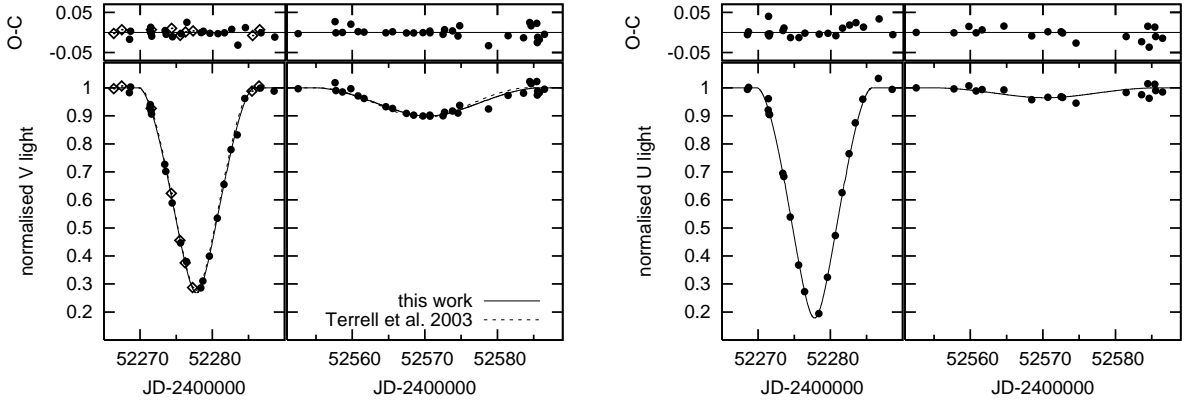}
\caption{\small{{\textit {V}} (left) and {\textit{U}} (right) light curves of OW Gem normalised
to 1 out of eclipse. The filled circles represent our measurements. The Derekas {\it et al}.$^7$
data are shown by diamonds. At the top the residuals (O-C) are placed, which demonstrate the quality
of the fit. The dashed line represents the model of Terrell {\it et al}.$^8$ where the secondary
minimum is shifted about 13 hours in respect to our model (solid line). }}
\label{VU_endcb}
\end{center}
\end{figure}

\begin{figure}[!b]
\begin{center}
\includegraphics[angle=270,width=0.75\textwidth]{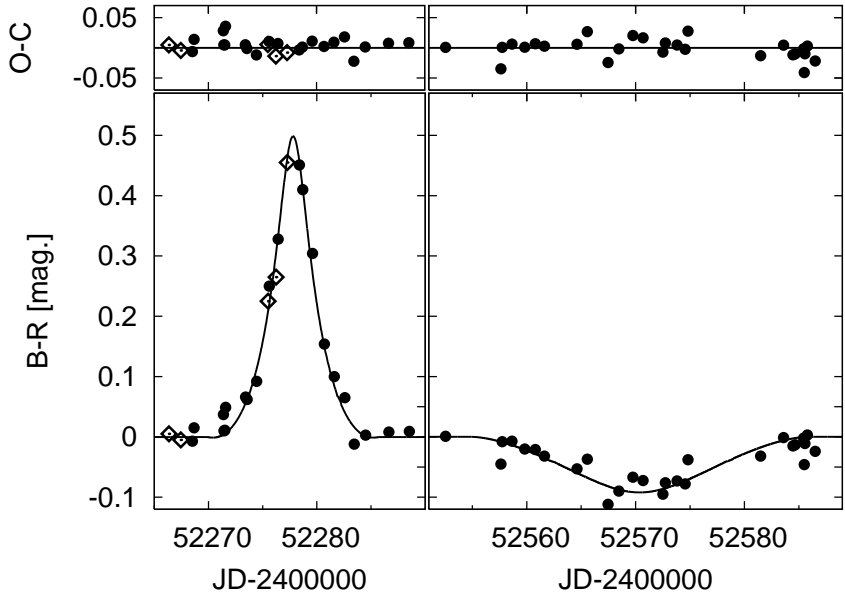}
\caption{\small{The {\textit{B-R}} color index during the primary and the secondary eclipses
of OW Gem. The filled circles represent our measurements. The Derekas~et~al.$^7$ data are shown
by diamonds. At the top, the residuals (O-C) are placed, which demonstrate the quality of the fit.}}
\label{B-R_endcb}
\end{center}
\end{figure}

The orbital inclination and radius of the primary component $R_1$ have been found by a search of the "whole"
$\chi^2$ surface for their possible values. The method relies on the execution of many fits where the two wanted
parameters are fixed and they are changed in the next runs with the assigned resolution. Such a map of the
$\chi^2$ function usually allows to find a global minimum, and so proper values of the parameters. The surface
$\chi^2(i,R_1)$ has been obtained by calculating a grid of $130$ values of $\chi^2$ (Figure \ref{SIEC3D_CHI_Wcb}),
for which a 3th order polynomial $f(x,y)$ was fitted. Later on, the minimum of the function $f(x,y)$ was found.
In this way, the inclination $i$ and the radius of primary component $R_1$ have been determined, and respondent
the values of the radius of secondary componenet $R_2$, temperature T$_2$ and the luminosity L$_1$.
The resulted value of the inclination is shown in the Table \ref{parorb}. Table \ref{parfot} presents our physical
parameters in comparison to those of other authors. Figure \ref{VU_endcb} demonstrates the quality of the fit to
{\it U} and {\it V} light curves, and Figure \ref{B-R_endcb} demonstrates the variations in the {\it B-R} color
index during the primary and the secondary eclipses of OW Gem.\\

\begin{table}[!h]
\begin{center}
\caption{\small{A comparison of the physical parameters obtained by us, Terrell {\it et al}.
and Griffin $\&$ Duquennoy.}}
\label{parfot}
{\small
\begin{tabular}{lllll}
                    &This work          & Terrell {\it et al}.$^8$ & G$\&$D$^5$    &unit\\

$T_1^{\star}$       & $7100$            & $7100$                   & $7100$        &$K$        \\
$T_2$               & $4975 \pm 20$     & $4917 \pm 110$           & $4800$        &$K$        \\
$\Omega_1$          & $33.34 \pm 0.21$  & $35.15 \pm 0.03$         & --            &--      \\
$\Omega_2$          & $24.17 \pm 0.15$  & $24.19 \pm 0.01$         & --            &--      \\
$R_1$               & $32.32 \pm 0.22$  & $30.9 \pm 0.3$           & $30 \pm 3$    &$R_{\odot}$\\
$R_2$               & $32.56 \pm 0.23$  & $31.7 \pm 0.3$           & $35 \pm 3$    &$R_{\odot}$\\
$M_1$               & $5.49 \pm 0.21$   & $5.8 \pm 0.2$            & $5.9 \pm 0.3$ &$M_{\odot}$\\ 
$M_2$               & $3.80 \pm 0.16$   & $3.9 \pm 0.1$            & $4.0 \pm 0.2$ &$M_{\odot}$\\ 
$(L_1/(L_1+L_2))_U$ & $0.949 \pm 0.011$ & $0.946 \pm 0.008$        & $0.945$       &--      \\
$(L_1/(L_1+L_2))_B$ & $0.921 \pm 0.007$ & $0.924 \pm 0.005$        & $0.899$       &--      \\
$(L_1/(L_1+L_2))_V$ & $0.851 \pm 0.007$ & $0.868 \pm 0.006$        & $0.834$       &--      \\
$(L_1/(L_1+L_2))_R$ & $0.803 \pm 0.008$ & $0.815 \pm 0.004$        & --            &--      \\
$(L_1/(L_1+L_2))_I$ & $0.757 \pm 0.009$ & $0.761 \pm 0.005$        & --            &--      \\
\multicolumn{5}{l}{$\star$ - {\it adopted}}
\end{tabular}
}
\end{center}
\end{table}

\noindent{\it OW Gem spectrum}\\

\begin{figure}[!h]
\begin{center}
\includegraphics[width=1.1\textwidth,angle=270]{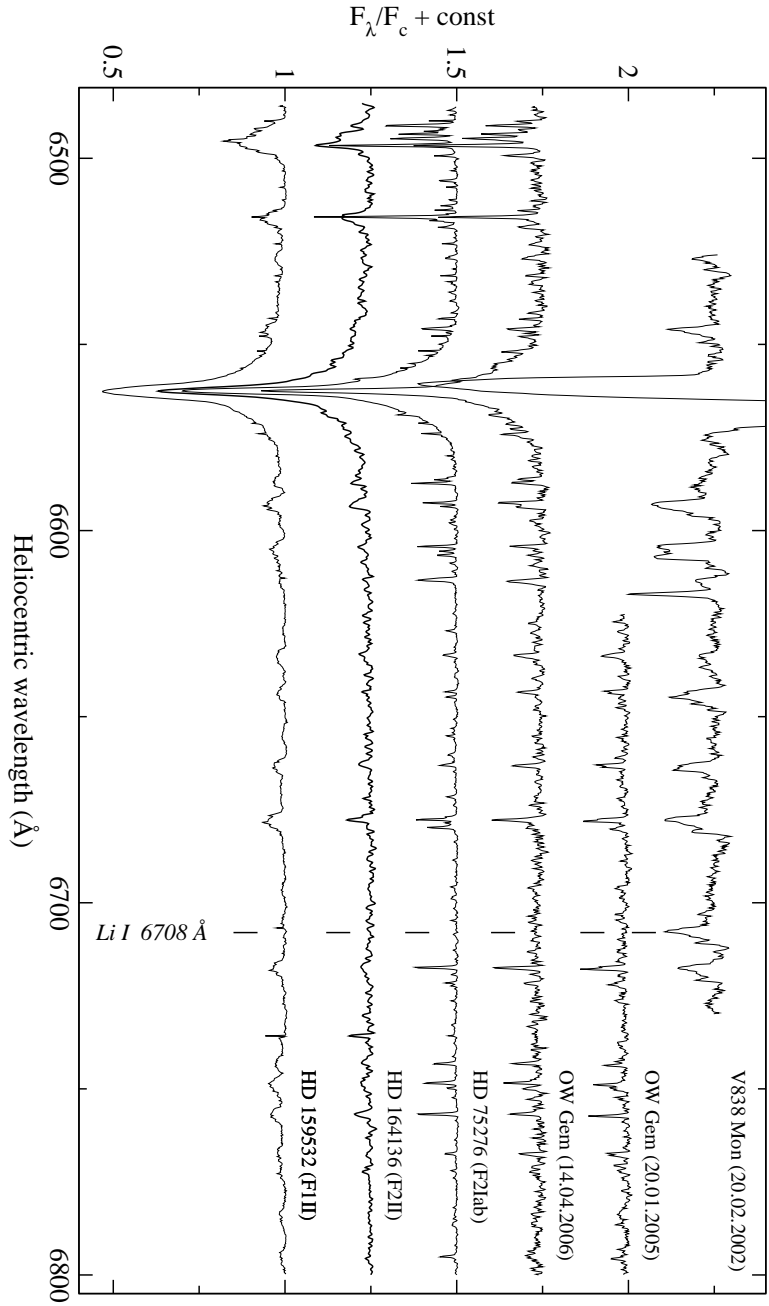}
\caption{\small{A comparison of our OW Gem spectra with the spectra of HD~75276, HD~164136 ($\nu$~Her),
HD~159532 and the possible merger V838 Mon. The spectra of the comparison stars are shifted to the
heliocentric wavelengths of the OW~Gem spectra. The spectra of HD~75276 and HD~159532 are degraded to
the OW~Gem spectra resolution. The H$_{\alpha}$ emission component in the spectrum of V838~Mon is truncated
for clarity.}}
\label{spectHer}
\end{center}
\end{figure}

We used the coude-spectrograph of the 2m RCC telescope at the Rozhen Observatory (Bulgaria) to obtain spectra
of OW Gem, with a resolving power R$\sim$15000, on January~20,~2005 ($\phi \sim$ 0.88) and April~14,~2006
($\phi \sim$ 0.24). The spectral regions covered were 6620-6825~$\AA$ and 6470-6820~$\AA$ respectively.
In Figure \ref{spectHer} the spectra of OW Gem are compared with spectra of HD~164136~($F2II$),
HD~75276~($F2Iab$) and HD~159532~($F1II$). The spectrum of HD~164136~($\nu$ Her) is from the Indo-U.S.
library of coud\'e feed stellar spectra (Valdes {\it et al}.$^{24}$) and the spectra of HD~75276 and HD~159532 are
from the UVES library of high-resolution spectra (Bagnulo {\it et al}.$^{25}$). Additionally, a spectrum of the
possible merger V838 Mon (Tylenda $\&$ Soker$^{26}$, and references therein), obtained at the Rozhen
Observatory with the same resolution as the OW Gem spectra, is shown in Figure \ref{spectHer} as well.

In the both spectra the radial velocities were measured (Table \ref{radTor}). In the spectrum on
April 14, 2006, obtained during the secondary eclipse, we measured the radial velocity of the primary component only.

Griffin $\&$ Duquennoy$^5$ classified the primary component of OW~Gem as an $F2Ib-II$ star. We do not have enough
spectral observations and intention to make a detailed spectral classification of both components. However,
in Figure~\ref{spectHer} it is obvious that the lines in the OW~Gem spectrum on April~14,~2006 dominated by the
primary component are very similar to the ones in the $F2Iab$ spectrum of HD~75276. The most remarkable difference
between the spectrum of the OW Gem primary and the spectra of $\nu$~Her (noted by Griffin~$\&$~Duquennoy$^5$)
and HD~159532 is the rotational velocity which is $V\sin{i}=28 km s^{-1}$ for $\nu$~Her and $V\sin{i}=105 km s^{-1}$
for HD~159532 (Snow {\it et al}.$^{27}$). Based on the January~20,~2005 spectrum only, we cannot say anything about
the secondary companion spectral class.

The above spectral regions were chosen with the aim of checking up the presence of a weak Li~I~6708~$\AA$ line
in the spectrum of the secondary component suggested by Griffin $\&$ Duquennoy$^5$. These authors measured an
equivalent width of about $10$ or $15 m\AA$ for this lithium line in the composite spectrum. Our April~14,~2006
spectrum (just during the secondary eclipse) is dominated by the primary component. If the lithium line is
present then it should be detectable in the January~20,~2005 spectrum (about 5 months before the primary
eclipse) in which most of the absorptions are double because of the blending of both component lines. The
quality of our spectra is good enough to identify and measure such weak absorptions. As can be seen in
Figure~\ref{spectHer}, in both spectra there are several faint features with equivalent widths of the order of
10-15~$m\AA$ in the vicinity of the lithium line. On one hand, none of these faint features disappear in the
spectrum on April~14,~2006, as we would expect if the weak lithium line were present only in the secondary
spectrum. On the other hand, the radial velocity measurements show that all of these faint features are far
from the expected lithium line position for both stars. Therefore, we cannot confirm the presence of the
Li~I~6708~$\AA$ absorption line, neither in the primary component spectrum nor in the secondary component
spectrum.

\newpage

To explain the unusual evolutionary status of the components, Eggleton$^{28}$ suggested that OW Gem is a
former triple star in which the F supergiant is a merged remnant of a close sub-binary. He pointed out that
it is very difficult to confirm that a particular star is or is not the result of a merger. A merger remnant
could be an unusually rapidly rotating star (Eggleton$^{28}$) but it is obvious from Figure~\ref{spectHer}
that this is not the case of the OW~Gem primary component. After the merging, a relatively strong
Li~I~6708~$\AA$ absorption line is present in the spectrum of V838~Mon (Figure~\ref{spectHer}). As was
noted above, this line is missing in the spectrum of the OW~Gem primary component. Hence, we can consider
the slow rotation and the lithium line absence only as an indication that, in case the primary component in
OW~Gem is a merger remnant, the merger event took place a long time ago.\\

\noindent{\it Conclusions}\\

\begin{figure}[!t]
\begin{center}
\includegraphics[angle=270,width=1.0\textwidth]{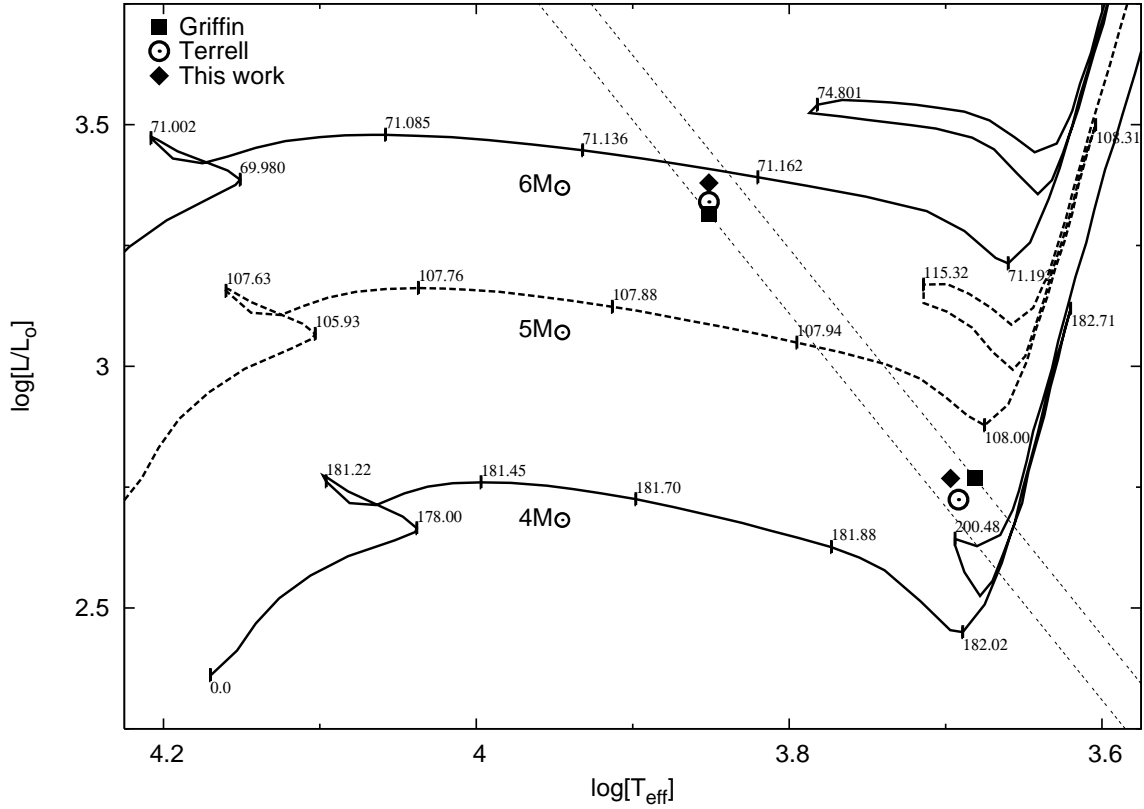}
\caption{\small{Both OW Gem components in confrontation with Girardi's {\it et al}.$^{28}$ solar
metallicity evolutionary tracks. The diagonal dashed lines are lines of constant radius for
30 $R_{\odot}$ and 35 $R_{\odot}$. The evolution time in millions of years is denoted by
short vertical lines with numerical value.}}
\label{HR}
\end{center}
\end{figure}

The full set of orbital and physical parameters for OW Gem have been obtained with independently collected
photometric data. A slightly better values pair's of parameters $\omega=140^{\circ}.73$ and $e=0.5286$
was obtained in our work in comparison to the previous analysis. Both the eccentricity and the periastron
argument calculated in this paper give a better fit to the observations, especially to the best timing of
the secondary minimum. This was possible by using new data including three secondary minima. Our results
underline the advantage of the simultaneous analysis of light and velocity curves. The new model have
supplied a better estimate of the radii of the OW~Gem components, using good quality multicolour photometry
as well as  more reliable temperature ratios of the stars. However, we were not able to significantly change
the values of the masses and the mass ratio of the components, confirming once again the unclear evolutionary
status of the system, in which two massive stars with considerably different masses ($\sim6 M_\odot$ and
$\sim4 M_\odot$) are placed in a very short stage of evolution of the supergiants. A confrontation with the
solar metallicity evolutionary tracks from Girardi~et~al.$^{29}$ is presented in Figure~\ref{HR}.
The less massive star is about 200 million years old. The more massive star is at least 100 million year
evolutionary younger. During this time it should finish its evolution as a supergiant. The current evolutionary
status of the system stands in contradiction with evolutionary models of the stars in binary systems and cannot
be explained either by loss or by transfer of mass (Griffin $\&$ Duquennoy$^5$). In order to explain the
observed parameters of OW~Gem, we should revise the stellar evolution theory. Another possibility is that
a merger took place. Eggleton$^{28}$ suggested that a triple system (close sub-binary $4 M_\odot$ + $2 M_\odot$
with short period about $2^d$ and the third component $4 M_\odot$ on wide orbit) can turn into a binary system.
The lack of lithium line detection in the present spectra is an indication that the merger event would have
had to have taken place long time ago. 

It seems that future attempts of modelling optical light and velocity curves will not result in significant
changes in our knowledge about the physical parameters of the system. Nevertheless, the OW~Geminorum case
still remains unexplained and an important case for understanding the evolution of the binary stars.
Especially the depths of the primary and secondary eclipses in deep UV and far infrared can give the
best, direct calibration of surface temperature for F and G supergiants. In present times it has became
possible to obtain an angular separation of such a few (about 3) milliarcseconds separated binary star
by optical interferometric observations (see e.g. Konacki~$\&$~Lane$^{30}$), and it gives opportunity
for verification of the distance to the OW Gem system.\\

\noindent{\it Acknowledgements}\\

We specially thank to Dr R.F. Griffin, who together with Dr A. Duquennoy have discovered unusally interesting
nature of OW Gem system, for his permission to use of the radial velocity data collected by them since their
paper in 1993. We are also deeply grateful to him for helpfull and kindness discussion.
We are very grateful to Dr B. Roukema for his language corrections and for N. Biernaczyk, S. Fr\c{a}ckowiak,
P. Oster, K. Rumi\'nski, E. \'Swierczy\'nski, M. Wi\c{e}cek, P. Wirkus, K. Wojtkowska, M. Wojtkowski,
P. Wychudzki for their contribution to collection of photometric data. This study was supported by
MNiSW grant No. N203 018 32$/$2338, grant UMK No. 340-A, and partly supported by the
Polish--Bulgarian Academy of Sciences exchange.\\


\centerline{\it References}
\noindent (1.) D. H. Kaiser, M. E. Baldwin, D. B.  Williams, D.B., {\it Inf. Bull. Var. Stars}, no. 3196, 1988.\\
~(2.) D. B. Williams, 1989, {\it J.A.A.V.S.O.}, {\bf{18}}, 7, 1989.\\
~(3.) D. H. Kaiser, {\it Inf. Bull. Var. Stars}, No. 3233, 1988.\\
~(4.) D. B. Williams, $\&$ D. H. Kaiser, {\it J.A.A.V.S.O.}, {\bf{20}}, 231, 1991.\\
~(5.) R. F. Griffin, $\&$ A. Duquennoy, {\it The Observatory}, {\bf{113}}, 53, 1993.\\
~(6.) R. F. Griffin, {\it The Observatory}, {\bf 113 }, 294, 1993.\\
~(7.) A. Derekas, {\it et al}., {\it Inf. Bull. Var. Stars}, no. 5239, 2002.\\
~(8.) D. Terrell, {\it et al}., {\it A.J.}, {\bf 126}, 902, 2003.\\
~(9.) D. H. Kaiser,{\it et al}., {\it Inf. Bull. Var. Stars}, no. 5347, 2002.\\
(10.) M. Miko{\l}ajewski, C. Ga{\l}an, D. Graczyk, {\it Inf. Bull. Var. Stars}, no. 5445, 2003\\
(11.) D. Terrell, D. H. Kaiser, D. B. Williams, {\it Inf. Bull. Var. Stars}, no. 4102, 1994.\\
(12.) K. K. Kwee, $\&$ H. Van Woerden, {\it Bull. Astron. Inst. Netherlands}, {\bf{12}}, 327, 1956.\\
(13.) T. Hager, {\it J.A.A.V.S.O.}, {\bf 24}, 9, 1996.\\
(14.) R. E. Wilson, $\&$ E. J. Devinney, {\it Ap.J.}, {\bf{166}}, 605, 1971.\\
(15.) R. E. Wilson, {\it Ap.J.}, {\bf{356}}, 613, 1990.\\
(16.) J. Kallrath, $\&$ E. F. Milone, {\it Eclipsing Binary Stars: Modeling and Analysis}, (New York: Springer), chap.4., 1998.\\
(17.) J. Kallrath, {\it et al}., {\it Ap.J.}, {\bf{508}}, 308, 1998.\\
(18.) R. E. Wilson, $\&$ P. Biermann, {\it A.$\&$A.}, {\bf{48}}, 349, 1976.\\
(19.) L. B. Lucy, 1967, {\it Z. Astrophys.}, {\bf{65}}, 89, 1967.\\
(20.) W. Van Hamme, {\it A.J.}, {\bf{106}}, 2096, 1993.\\
(21.) R. E. Wilson, {\it Ap.J.}, {\bf{234}}, 1054, 1979.\\
(22.) R. F. Griffin, {\it The Observatory}, {\bf{124}}, 136, 2004.\\
(23.) V. Strai\v{z}ys, $\&$ G. Kuriliene, {\it Ap.$\&$S.S.}, {\bf 80}, 353, 1981.\\
(24.) F. Valdes, {\it et al}., {\it Ap.J.S.}, {\bf{152}}, 251, 2004.\\
(25.) S. Bagnulo, {\it et al}., {\it The Messenger}, {\bf{114}}, 10, 2003.\\
(26.) R. Tylenda, $\&$ N. Soker, {\it A.$\&$A.}, {\bf{451}}, 223, 2006.\\
(27.) T. P. Snow, {\it et al}., {\it Ap.J.S.}, {\bf{95}}, 163, 1994.\\
(28.) P. P. Eggleton, in {\it Exotic Stars as Challenges to Evolution}, edited by A. C. Tout $\&$ W. Van Hamme,(A.S.P. Conf. Ser. 279, San Francisco), 2002, p. 37.\\
(29.) L. Girardi, {\it et al}., {\it A.$\&$A.}, {\bf{141}}, 371, 2000.\\
(30.) M. Konacki, $\&$ B. F. Lane, {\it Ap.J.}, {\bf{610}}, 443, 2004.\\


\newpage

{\it Figures:}

\newpage

\newpage

\newpage

\newpage

\newpage

\appendix{\it Appendix - photometric and radial velocity data:}

{\small
\begin{center}
\begin{longtable}{lrrrrr}

\caption{\normalsize { \it UBVRI} light curves of the 1995 primary and secondary eclipses collected
with EMI9558B photomultiplier.}
\label{photEMI}
\\
$JD-2400000$ & $\Delta U$ & $\Delta B$ & $\Delta V$ & $\Delta R$ & $\Delta I$ \\

\endfirsthead

$JD-2400000$ & $\Delta U$ & $\Delta B$ & $\Delta V$ & $\Delta R$ & $\Delta I$ \\

\endhead

\endfoot

\endlastfoot

$49723.4524$ &   $ 0.010$ &    $-0.320$ &    $-0.758$ &    $-1.026$ &    $-1.247$ \\
$49731.3870$ &   $-0.006$ &    $-0.340$ &    $-0.766$ &    $-1.020$ &    $-1.260$ \\
$49741.4052$ &   $-0.011$ &    $-0.320$ &    $-0.767$ &    $-1.022$ &    $-1.241$ \\
$49751.3617$ &   $ 0.004$ &    $-0.339$ &    $-0.743$ &    $-1.016$ &    $-1.232$ \\
$49751.5129$ &   $-0.004$ &    $-0.330$ &    $-0.765$ &    $-1.029$ &    $-1.233$ \\
$49754.2218$ &   $ 0.087$ &    $-0.233$ &    $-0.687$ &    $-0.954$ &    $-1.169$ \\
$49754.2680$ &   $ 0.041$ &    $-0.256$ &    $-0.695$ &    $-0.950$ &    $-1.185$ \\
$49754.3465$ &   $ 0.105$ &    $-0.243$ &    $-0.669$ &    $-0.938$ &    $-1.136$ \\
$49754.4260$ &   $ 0.107$ &    $-0.209$ &    $-0.655$ &    $-0.942$ &    $-1.150$ \\
$49756.2605$ &   $ 0.393$ &    $ 0.069$ &    $-0.416$ &    $-0.681$ &    $-0.937$ \\
$49756.4032$ &   $ 0.411$ &    $ 0.084$ &    $-0.378$ &    $-0.662$ &    $-0.898$ \\
$49757.2831$ &   $ 0.669$ &    $ 0.299$ &    $-0.188$ &    $-0.477$ &    $-0.774$ \\
$49758.4480$ &   $ 1.084$ &    $ 0.702$ &    $ 0.114$ &    $-0.232$ &    $-0.488$ \\
$49759.2654$ &   $ 1.410$ &    $ 0.911$ &    $ 0.290$ &    $-0.101$ &       --    \\
$49761.2366$ &   $ 1.777$ &    $ 1.336$ &    $ 0.597$ &    $ 0.201$ &    $-0.106$ \\
$49761.5347$ &      --    &    $ 1.191$ &    $ 0.507$ &    $ 0.097$ &    $-0.179$ \\
$49762.4387$ &   $ 1.221$ &    $ 0.851$ &    $ 0.235$ &    $-0.137$ &    $-0.412$ \\
$49763.5336$ &   $ 0.811$ &    $ 0.430$ &    $-0.082$ &    $-0.408$ &    $-0.682$ \\
$49764.4510$ &   $ 0.507$ &    $ 0.185$ &    $-0.303$ &    $-0.599$ &    $-0.847$ \\
$49765.4287$ &   $ 0.289$ &    $-0.033$ &    $-0.492$ &    $-0.782$ &    $-1.009$ \\
$49766.2876$ &   $ 0.143$ &    $-0.174$ &    $-0.563$ &    $-0.846$ &    $-1.089$ \\
$49767.3374$ &   $ 0.043$ &    $-0.279$ &    $-0.720$ &    $-0.966$ &    $-1.174$ \\
$49769.4922$ &   $-0.038$ &    $-0.315$ &    $-0.761$ &    $-1.007$ &    $-1.214$ \\
$49771.3802$ &   $ 0.004$ &    $-0.320$ &    $-0.749$ &    $-1.013$ &    $-1.221$ \\
$50040.4597$ &      --    &    $-0.377$ &    $-0.782$ &    $-1.016$ &    $-1.229$ \\
$50044.4597$ &   $ 0.005$ &    $-0.311$ &    $-0.720$ &    $-0.963$ &    $-1.171$ \\
$50047.4604$ &   $ 0.006$ &    $-0.296$ &    $-0.686$ &    $-0.927$ &    $-1.151$ \\
$50048.4013$ &      --    &    $-0.266$ &    $-0.679$ &    $-0.913$ &    $-1.086$ \\
$50066.4347$ &   $ 0.025$ &    $-0.306$ &    $-0.741$ &    $-0.989$ &    $-1.232$ \\
$50068.3413$ &      --    &    $-0.331$ &    $-0.734$ &    $-0.969$ &    $-1.215$ \\
$50884.4409$ &   $-0.044$ &    $-0.355$ &    $-0.750$ &    $-1.028$ &    $-1.231$ \\
$50895.3600$ &   $-0.040$ &    $-0.329$ &    $-0.764$ &    $-1.016$ &    $-1.259$ \\
$51078.5412$ &   $-0.015$ &    $-0.311$ &    $-0.760$ &    $-1.022$ &    $-1.262$ \\
$51196.3584$ &   $-0.025$ &    $-0.335$ &    $-0.756$ &    $-1.020$ &    $-1.239$ \\
$51257.4235$ &   $-0.008$ &    $-0.350$ &    $-0.768$ &    $-1.028$ &    $-1.269$ \\
\end{longtable}
\end{center}
}

{\small
\begin{center}
\begin{longtable}{lrrrrrrr}

\caption{\normalsize {\it UBV(RI)$_C$hc} light curves of the 2002 secondary eclipse collected
with Burle C31034 photomultiplier.}
\label{photBurle}
\\

$JD-2400000$ & $\Delta U$ & $\Delta B$ & $\Delta V$ & $\Delta R_C$ & $\Delta I_C$ & $\Delta c$ & $\Delta h$ \\

\endfirsthead

$JD-2400000$ & $\Delta U$ & $\Delta B$ & $\Delta V$ & $\Delta R_C$ & $\Delta I_C$ & $\Delta c$ & $\Delta h$ \\

\endhead

\endfoot

\endlastfoot

$52520.5887$ &    $-0.011$ &    $-0.399$ &    $-0.750$ &    $-1.089$ &    $-1.405$ & $-0.555$ & $-0.605$ \\
$52526.5821$ &    $-0.002$ &    $-0.401$ &    $-0.779$ &    $-1.114$ &    $-1.398$ & $-0.549$ & $-0.612$ \\
$52528.5610$ &    $ 0.022$ &    $-0.395$ &    $-0.763$ &    $-1.120$ &    $-1.423$ & $-0.569$ & $-0.574$ \\
$52530.5813$ &    $ 0.055$ &    $-0.406$ &    $-0.761$ &    $-1.090$ &    $-1.397$ & $-0.521$ & $-0.626$ \\
$52535.5425$ &    $ 0.031$ &    $-0.412$ &    $-0.772$ &    $-1.100$ &    $-1.400$ & $-0.560$ & $-0.578$ \\
$52537.5415$ &    $ 0.044$ &    $-0.382$ &    $-0.745$ &    $-1.084$ &    $-1.395$ & $-0.549$ & $-0.562$ \\
$52542.6051$ &    $ 0.044$ &    $-0.379$ &    $-0.748$ &    $-1.087$ &    $-1.376$ & $-0.560$ & $-0.565$ \\
$52550.5663$ &    $ 0.031$ &    $-0.388$ &    $-0.765$ &    $-1.088$ &    $-1.382$ & $-0.564$ & $-0.589$ \\
$52552.5397$ &    $ 0.034$ &    $-0.386$ &    $-0.751$ &    $-1.095$ &    $-1.384$ & $-0.562$ & $-0.596$ \\
$52558.6378$ &       --    &    $-0.353$ &    $-0.738$ &    $-1.054$ &    $-1.351$ & $-0.530$ & $-0.576$ \\
$52567.4629$ &       --    &    $-0.335$ &    $-0.650$ &    $-0.926$ &    $-1.205$ & $-0.475$ & $-0.505$ \\
$52568.4644$ &    $ 0.080$ &    $-0.325$ &    $-0.643$ &    $-0.939$ &    $-1.220$ & $-0.442$ & $-0.510$ \\
$52572.5023$ &    $ 0.068$ &    $-0.331$ &    $-0.640$ &    $-0.940$ &    $-1.217$ & $-0.441$ & $-0.490$ \\
$52574.5563$ &    $ 0.094$ &    $-0.307$ &    $-0.652$ &    $-0.934$ &    $-1.257$ & $-0.432$ & $-0.431$ \\
$52581.4831$ &    $ 0.052$ &    $-0.383$ &    $-0.724$ &    $-1.057$ &    $-1.351$ & $-0.511$ & $-0.577$ \\
$52584.4530$ &    $ 0.018$ &    $-0.420$ &    $-0.778$ &    $-1.112$ &    $-1.406$ & $-0.555$ & $-0.605$ \\
$52584.6540$ &    $ 0.074$ &    $-0.391$ &    $-0.770$ &    $-1.086$ &    $-1.390$ & $-0.571$ & $-0.579$ \\
$52585.4333$ &    $ 0.020$ &    $-0.401$ &    $-0.778$ &    $-1.106$ &    $-1.408$ & $-0.532$ & $-0.578$ \\
$52585.5563$ &    $ 0.045$ &    $-0.388$ &    $-0.741$ &    $-1.084$ &    $-1.386$ & $-0.524$ & $-0.574$ \\
$52586.5032$ &    $ 0.050$ &    $-0.392$ &    $-0.749$ &    $-1.076$ &    $-1.386$ & $-0.549$ & $-0.543$ \\
$52603.4456$ &       --    &    $-0.364$ &    $-0.754$ &    $-1.115$ &    $-1.406$ & $-0.534$ & $-0.586$ \\
$52615.4383$ &    $ 0.031$ &    $-0.383$ &    $-0.750$ &    $-1.091$ &    $-1.378$ & $-0.547$ & $-0.586$ \\
$52618.5081$ &    $ 0.054$ &    $-0.389$ &    $-0.751$ &    $-1.107$ &    $-1.408$ & $-0.547$ & $-0.579$ \\
$52714.3303$ &    $ 0.021$ &    $-0.388$ &    $-0.744$ &    $-1.097$ &    $-1.391$ & $-0.584$ & $-0.586$ \\
$52889.6166$ &       --    &    $-0.357$ &    $-0.713$ &    $-1.068$ &    $-1.365$ & $-0.533$ & $-0.522$ \\
$52890.6166$ &    $ 0.027$ &    $-0.367$ &    $-0.753$ &    $-1.068$ &    $-1.361$ & $-0.502$ & $-0.552$ \\
$52904.5900$ &    $ 0.038$ &    $-0.398$ &    $-0.769$ &    $-1.096$ &    $-1.433$ & $-0.517$ & $-0.597$ \\
$52922.5165$ &    $ 0.043$ &    $-0.382$ &    $-0.736$ &    $-1.086$ &    $-1.379$ & $-0.559$ & $-0.604$ \\
$52935.5517$ &    $ 0.049$ &    $-0.384$ &    $-0.744$ &    $-1.083$ &    $-1.382$ & $-0.541$ & $-0.594$ \\
$52950.4901$ &       --    &    $-0.423$ &    $-0.763$ &    $-1.142$ &    $-1.385$ & $-0.596$ & $-0.577$ \\
$52985.3382$ &    $ 0.032$ &    $-0.378$ &    $-0.771$ &    $-1.101$ &    $-1.409$ &--&--\\
$53008.5002$ &       --    &    $-0.378$ &    $-0.714$ &    $-1.086$ &    $-1.351$ &--&--\\
$53026.5686$ &       --    &    $-0.421$ &    $-0.764$ &    $-1.125$ &       --    &--&--\\
$53056.3132$ &    $ 0.051$ &    $-0.352$ &    $-0.738$ &    $-1.090$ &    $-1.381$ &--&--\\
$53069.4435$ &    $ 0.059$ &    $-0.397$ &    $-0.749$ &    $-1.114$ &    $-1.373$ &--&--\\
$53077.4511$ &    $ 0.051$ &    $-0.417$ &    $-0.769$ &    $-1.121$ &    $-1.383$ &--&--\\
$53094.3720$ &    $ 0.009$ &    $-0.349$ &    $-0.779$ &    $-1.103$ &    $-1.403$ &--&--\\
$53110.3685$ &    $ 0.029$ &    $-0.367$ &    $-0.758$ &    $-1.093$ &    $-1.428$ &--&--\\
$53122.3420$ &       --    &    $-0.409$ &    $-0.760$ &    $-1.087$ &    $-1.344$ &--&--\\
$53273.6200$ &       --    &    $-0.390$ &    $-0.741$ &    $-1.098$ &    $-1.382$ &--&--\\
$53291.5840$ &    $ 0.015$ &    $-0.386$ &    $-0.771$ &    $-1.118$ &    $-1.407$ &--&--\\
\end{longtable}
\end{center}
}

{\small
\begin{center}
\begin{longtable}{lrrrrrrrrrr}

\caption{\normalsize {\it UBV(RI)$_C$} data obtained with the SBIG STL~11000 CCD camera. The columns
with HJD+ denote the fraction of the day.}
\label{photSTL11000}
\\
$HJD$&HJD+&$\Delta U$&$HJD+$&$\Delta B$&$HJD+$&$\Delta V$&$HJD+$&$\Delta R_C$&$HJD+$&$\Delta I_C$\\

\endfirsthead

$HJD$&HJD+&$\Delta U$&$HJD+$&$\Delta B$&$HJD+$&$\Delta V$&$HJD+$&$\Delta R_C$&$HJD+$&$\Delta I_C$\\

\endhead

\endfoot

\endlastfoot

$2453449$&$.4107$&$-0.131$&$.4380$&$-0.373$&   --  &   --   &   --  &   --   &   --  &   --   \\
$2453463$&$.3898$&$-0.128$&$.3955$&$-0.378$&$.4031$&$-0.716$&$.4066$&$-0.993$&$.4107$&$-1.329$\\
$2453477$&   --  &   --   &$.3587$&$-0.389$&$.3556$&$-0.725$&$.3684$&$-1.019$&$.3709$&$-1.339$\\
\end{longtable}
\end{center}
}

{\small
\begin{center}
\begin{longtable}{lrrrrrrrrrr}

\caption{\normalsize {\it UBV(RI)$_C$} data of the 2006 secondary eclipse obtained with the SBIG STL~1001 CCD camera.
The columns with HJD+ denote the fraction of the day.}
\label{photSTL1001}
\\
$HJD$&$HJD+$&$\Delta U$&$HJD+$&$\Delta B$&$HJD+$&$\Delta V$&$HJD+$&$\Delta R_C$&$HJD+$&$\Delta I_C$\\

\endfirsthead

$HJD$&$HJD+$&$\Delta U$&$HJD+$&$\Delta B$&$HJD+$&$\Delta V$&$HJD+$&$\Delta R_C$&$HJD+$&$\Delta I_C$\\

\endhead

\endfoot

\endlastfoot

$2453648$&$.6597$&$-0.110$&$.6614$&$-0.381$&$.6622$&$-0.748$&$.6638$&$-1.032$&$.6644$&$-1.375$\\
$2453745$&$.6161$&$-0.152$&$.6129$&$-0.388$&$.6104$&$-0.750$&$.6185$&$-1.036$&$.6196$&$-1.379$\\
$2453760$&   --  &   --   &$.4652$&$-0.388$&$.4591$&$-0.751$&$.4784$&$-1.028$&$.4805$&$-1.382$\\
$2453789$&$.2514$&$-0.116$&$.2476$&$-0.381$&$.2451$&$-0.750$&$.2551$&$-1.027$&$.2582$&$-1.389$\\
$2453799$&$.3832$&$-0.137$&$.3861$&$-0.385$&$.3873$&$-0.743$&$.3887$&$-1.024$&$.3894$&$-1.366$\\
$2453801$&$.3573$&$-0.135$&$.3602$&$-0.376$&$.3612$&$-0.744$&$.3621$&$-1.022$&$.3630$&$-1.374$\\
$2453816$&$.3272$&$-0.135$&$.3267$&$-0.389$&$.3348$&$-0.746$&$.3307$&$-1.016$&   --  &   --   \\
$2453818$&$.3875$&$-0.147$&$.3889$&$-0.390$&$.3893$&$-0.754$&$.3898$&$-1.005$&$.3901$&$-1.337$\\
$2453819$&$.3655$&$-0.127$&$.3683$&$-0.368$&$.3700$&$-0.725$&$.3713$&$-0.983$&$.3723$&$-1.321$\\
$2453828$&   --  &   --   &$.3327$&$-0.328$&$.3281$&$-0.643$&$.3465$&$-0.896$&$.3492$&$-1.203$\\
$2453829$&$.2706$&$-0.103$&$.2652$&$-0.323$&$.2613$&$-0.647$&$.2741$&$-0.889$&$.2908$&$-1.194$\\
$2453829$&   --  &   --   &   --  &   --   &$.2931$&$-0.642$&$.2898$&$-0.884$&   --  &   --   \\
$2453831$&$.3043$&$-0.102$&$.3072$&$-0.335$&$.3007$&$-0.659$&$.3081$&$-0.894$&$.3089$&$-1.208$\\
$2453832$&   --  &   --   &$.3913$&$-0.342$&$.3922$&$-0.664$&$.3928$&$-0.904$&$.3933$&$-1.231$\\
$2453833$&   --  &   --   &$.3907$&$-0.350$&$.3796$&$-0.688$&$.3808$&$-0.948$&$.3819$&$-1.250$\\
$2453837$&   --  &   --   &   --  &   --   &$.3607$&$-0.673$&   --  &   --   &   --  &   --   \\
$2453844$&   --  &   --   &$.3568$&$-0.394$&$.3523$&$-0.738$&$.3603$&$-1.032$&$.3617$&$-1.368$\\
$2454025$&$.4877$&$-0.135$&$.4827$&$-0.386$&$.4812$&$-0.741$&$.4913$&$-1.035$&$.4922$&$-1.368$\\
$2454066$&$.4471$&$-0.144$&$.4389$&$-0.373$&$.4354$&$-0.738$&$.4406$&$-1.029$&$.4418$&$-1.376$\\
$2454120$&$.6090$&$-0.169$&$.6044$&$-0.425$&$.6031$&$-0.794$&$.6066$&$-1.043$&$.6075$&$-1.376$\\
$2454128$&$.3383$&$-0.136$&$.3435$&$-0.421$&$.3310$&$-0.770$&$.3469$&$-1.030$&$.3496$&$-1.371$\\
$2454188$&   --  &   --   &$.2808$&$-0.431$&$.2827$&$-0.773$&$.2847$&$-1.047$&$.2882$&$-1.383$\\
$2454207$&$.3560$&$-0.157$&$.3519$&$-0.449$&$.3497$&$-0.795$&$.3483$&$-1.047$&$.3466$&$-1.382$\\
\end{longtable}
\end{center}
}

{\small
\begin{center}
\begin{longtable}{llllll}

\caption{\normalsize Radial velocity data obtained by R.F. Griffin and A. Duquennoy.}
\label{rvGD}
\\
Hel. Date       & HJD$-2400000$& \multicolumn{2}{c}{Velocity $[km s^{-1}]$} & Source$^{\star}$ & Weight\\
                &              & prim. & sec. & & \\
\endfirsthead

Hel. Date       & HJD$-2400000$& \multicolumn{2}{c}{Velocity} $[km s^{-1}]$& Source$^{\star}$ & Weight\\
                &              & prim. & sec. & & \\
\endhead

\endfoot

\endlastfoot

1988 Nov. ~3.21 &  $47468.71$    &$-19.0$   &  --  &    OHP & $1$\\
1988 Nov. ~7.19 &  $47472.69$    &$-18.4$   &  --  &    OHP & $1$\\
1988 Dec. ~6.09 &  $47501.59$    &$-17.0$   &  --  & Cambridge-old & $1/8$\\
1988 Dec. 13.04 &  $47508.54$    &$-16.5$   &  --  & Cambridge-old & $1/8$\\
1988 Dec. 20.04 &  $47515.54$    &$-14.0$   &  --  & Cambridge-old & $1/8$\\
1989 Jan. ~5.03 &  $47531.53$    &$-15.3$   &  --  & Cambridge-old & $1/8$\\
1989 Jan. 17.98 &  $47544.48$    &$-12.5$   &  --  & Cambridge-old & $1/8$\\
1989 Feb. 24.18 &  $47581.68$    &$-10.8$   & $+6.1$ &    ESO & $1$\\
1989 Mar. 26.84 &  $47612.34$    &$ -8.9$   & $+0.4$ &    OHP & $1$\\
1989 Apr. 29.82 &  $47646.32$    &$ -7.0$   & $+0.6$ &    OHP & $1$\\
1989 Oct. 30.16 &  $47829.66$    &$ -0.7$   &$-14.9$ &    OHP & $1$\\
1989 Nov. 17.14 &  $47847.64$    &$ +2.6$   &  --  & Cambridge-old & $1/8$\\
1989 Dec. 23.04 &  $47883.54$    &$ +1.3$   &  --  & Cambridge-old & $1/8$\\
1990 Jan. 14.01 &  $47905.51$    &$ +2.2$   &  --  & Cambridge-old & $1/8$\\
1990 Jan. 30.02 &  $47921.52$    &$ +2.1$   &$-18.4$ &    OHP & $1$\\
1990 Feb. 12.14 &  $47934.64$    &$ +1.5$   &$-18.9$ &    ESO & $1$\\
1990 Apr. ~4.84 &  $47986.34$    &$ +3.2$   &  --  & Cambridge-old & $1/8$\\
1990 Oct. ~7.20 &  $48171.70$    &$ +6.6$   &  --  & Cambridge-old & $1/8$\\
1990 Dec. ~4.13 &  $48229.63$    &$ +7.2$   &  --  & Cambridge-old & $1/8$\\
1991 Jan. 26.01 &  $48282.51$    &$ +6.6$   &$-22.6$ &    OHP & $1$\\
1991 Mar. 13.85 &  $48329.35$    &$ +6.5$   &  --  & Cambridge-old & $1/8$\\
1991 Apr. ~3.87 &  $48350.37$    &$ +3.7$   &  --  & Cambridge-old & $1/8$\\
1991 Oct. 29.16 &  $48558.66$    &$-30.4$   &$+30.2$ &    OHP & $2$\\
1991 Dec. 19.04 &  $48609.54$    &$-32.5$   &$+33.5$ &    OHP & $2$\\
1992 Jan. 14.01 &  $48635.51$    &$-29.9$   &$+30.2$ &    OHP & $2$\\
1992 Jan. 18.00 &  $48639.50$    &$-29.1$   &$+30.3$ &    OHP & $2$\\
1992 Jan. 24.06 &  $48645.56$    &$-28.4$   &$+31.0$ &    OHP & $2$\\
1992 Feb. 27.30 &  $48679.80$    &$-24.1$   &$+22.3$ &    DAO & $2$\\
1992 Apr. 22.85 &  $48735.35$    &$-18.5$   &$+14.0$ &    OHP & $1$\\
1992 Aug. 16.13 &  $48850.63$    &\multicolumn{2}{c}{$-9.0$}     &    OHP & $0$\\
1992 Aug. 17.13 &  $48851.63$    &\multicolumn{2}{c}{$-9.1$}     &    OHP & $0$\\
1992 Dec. 18.07 &  $48974.57$    &\multicolumn{2}{c}{$-4.8$}     &    OHP & $0$\\
1993 Feb. 15.93 &  $49034.43$    &\multicolumn{2}{c}{$-3.4$}     &    OHP & $0$\\
1993 Mar. 22.86 &  $49069.36$    & $-1.4$   &$-16.8$ &    OHP & $1$\\
1993 Apr. 20.83 &  $49098.33$    & $-0.4$   &$-15.5$ &    OHP & $1$\\
1993 Aug. 30.16 &  $49229.66$    &\multicolumn{2}{c}{$+1.5$}     &    OHP & $0$\\
1993 Sep. 12.14 &  $49242.64$    & $+4.2$   &$-10.6$ &    OHP & $1$\\
1993 Nov. ~6.34 &  $49297.84$    &\multicolumn{2}{c}{$+3.3$}     &    OHP & $0$\\
1994 Jan. ~3.05 &  $49355.55$    & $+6.4$   &  --  &    OHP & $1$\\
1994 Feb. 19.88 &  $49403.38$    & $+6.1$   &$-18.7$ &    OHP & $1$\\
1994 Apr. 30.84 &  $49473.34$    &\multicolumn{2}{c}{$+5.7$}     &    OHP & $0$\\
1994 Dec. 12.13 &  $49698.63$    & $-1.6$   &$-17.0$ &    OHP & $1$\\
1995 Jan. ~5.06 &  $49722.56$    &\multicolumn{2}{c}{$-4.4$}     &    OHP & $0$\\
1996 Jan. ~1.08 &  $50083.58$    &\multicolumn{2}{c}{$-10.1$}    &    OHP & $0$\\
1996 Mar. 31.87 &  $50174.37$    &\multicolumn{2}{c}{$-6.4$}     &    OHP & $0$\\
1996 Dec. 16.07 &  $50433.57$    &\multicolumn{2}{c}{$+1.5$}     &    OHP & $0$\\
1997 Jan. 26.03 &  $50474.53$    & $+2.8$   &$-15.0$ &    OHP & $1$\\
1997 Sep. 11.16 &  $50702.66$    &\multicolumn{2}{c}{$+5.2$}     &    OHP & $0$\\
1997 Dec. 21.06 &  $50803.56$    & $+6.5$   &$-15.9$ &    OHP & $1$\\
2000 Jan. ~9.02 &  $51552.52$    & $-1.5$   &$ -9.5$ & Cambridge & $1$\\
2000 Feb. 28.86 &  $51603.36$    & $-0.7$   &$-15.9$ & Cambridge & $1$\\
2000 Apr. ~6.85 &  $51641.35$    & $+1.3$   &$-11.5$ & Cambridge & $1$\\
2000 Nov. 13.17 &  $51861.67$    & $+5.4$   &$-15.4$ & Cambridge & $1$\\
2001 Jan. ~7.02 &  $51916.52$    & $+5.6$   &$-18.0$ & Cambridge & $1$\\
2001 Feb. 13.93 &  $51954.43$    & $+6.3$   &$-20.0$ & Cambridge & $1$\\
2001 Nov. 14.20 &  $52227.70$    & $-2.7$   &$-13.8$ & Cambridge & $1$\\
2001 Dec. 12.11 &  $52255.61$    &\multicolumn{2}{c}{$-5.9$}     & Cambridge & $0$\\
2001 Dec. 22.05 &  $52265.55$    &$ -9.2$   &$ -1.7$ & Cambridge & $1$\\
2002 Jan. 17.99 &  $52292.49$    &$-17.7$   &$+18.2$ & Cambridge & $1$\\
2002 Jan. 24.97 &  $52299.47$    &$-20.2$   &$+17.2$ & Cambridge & $1$\\
2002 Feb. ~5.98 &  $52311.48$    &$-23.8$   &$+26.0$ & Cambridge & $2$\\
2002 Feb. 16.90 &  $52322.40$    &$-27.5$   &$+25.7$ & Cambridge & $2$\\
2002 Mar. ~1.88 &  $52335.38$    &$-29.5$   &$+31.2$ & Cambridge & $2$\\
2002 Mar. ~9.90 &  $52343.40$    &$-31.3$   &$+29.6$ & Cambridge & $2$\\
2002 Mar. 27.87 &  $52361.37$    &$-33.1$   &$+34.7$ & Cambridge & $2$\\
2002 Apr. ~6.86 &  $52371.36$    &$-33.0$   &$+34.4$ & Cambridge & $2$\\
2002 Apr. 23.85 &  $52388.35$    &$-32.2$   &$+35.2$ & Cambridge & $2$\\
2002 Oct. 24.18 &  $52571.68$    &$-13.2$   &  --  & Cambridge & $1$\\
2002 Nov. ~7.20 &  $52585.70$    &$-12.3$   &  --  & Cambridge & $1$\\
2003 Jan. ~6.01 &  $52645.51$    &$ -8.3$   &$ -4.1$ & Cambridge & $1$\\
2003 Dec. 15.13 &  $52988.63$    &$ +2.8$   &$-14.6$ & Cambridge & $1$\\
2004 Feb. 27.94 &  $53063.44$    &$ +4.0$   &$-18.2$ & Cambridge & $1$\\
2005 Jan. 11.03 &  $53381.53$    &$ +5.6$   &$-18.6$ & Cambridge & $1$\\
2005 Nov. 25.16 &  $53699.66$    &$-26.4$   &$+27.9$ & Cambridge & $2$\\
2005 Dec. 28.09 &  $53732.59$    &$-22.4$   &$+23.5$ & Cambridge & $2$\\
2006 Feb. 20.93 &  $53787.43$    &$-17.0$   &$ +9.3$ & Cambridge & $1$\\
2007 Apr. 10.87 &  $54201.37$    &$ +1.5$   &$-12.2$ & Cambridge & $1$\\

\end{longtable}
\end{center}
}

{\small \noindent Description for the table \ref{rvGD}:\\

\noindent $^{\star}$Sources:\\
{\it HPO} - Coravel at Haute-Provence Observatory,\\
{\it Cambridge-old} - the old original radial-velocity spectrometer at Cambridge, with which Griffin first
developed the cross-correlation method of measuring velocities (ApJ 148, 465, 1967),\\
{\it Cambridge} - Coravel instrument currently working at Cambridge Observatory,\\
{\it DAO} - instrument at Dominion Astrophysical Observatory,\\
{\it ESO} - another spectrometer similar Coravel.\\

The velocities written between the columns for the primary and secondary have been reduced as if the system
were single-lined and the zero weight were applied in those cases.\\
The data before 1992 Aug. 16 were published already by Griffin $\&$ Duquennoy
in 1993, but now they have been corrected by Dr Griffin and are presented here once again.\\
}

{\small
\begin{center}
\begin{longtable}{llllc}

\caption{\normalsize Radial velocity data obtained from Rozhen Observatory spectra.}
\label{radTor}
\\
Date         & JD$-2400000$ & \multicolumn{2}{c}{Velocity $[km s^{-1}]$} & Weight \\
             &              & prim. & ~sec. & \\

\endfirsthead

Date         & JD$-2400000$ & \multicolumn{2}{c}{Velocity $[km s^{-1}]$} & Weight \\
             &              & prim. & ~sec. & \\

\endhead

\endfoot

\endlastfoot

2005 Jan. 20 & $53391.31$   & ~$+5.5 \pm 0.8$  & $-18.2 \pm 1.3$ & $1$ \\
2006 Apr. 14 & $52581.70$   &$-12.5^{\star} \pm 0.9$ & & $1$ \\
\multicolumn{5}{l}{$^{\star}$ this is measurement of the blend of primary and secondary but the weight one}\\
\multicolumn{5}{l}{was applied for calculations}\\
\end{longtable}
\end{center}
}

\end{document}